\documentstyle[preprint,aps,version2]{revtex}  
\begin{document}
\thispagestyle{empty}

\renewcommand{\small}{\normalsize} 

\font\fortssbx=cmssbx10 scaled \magstep2
\hbox to \hsize{
\hskip.5in \raise.1in\hbox{\fortssbx University of Wisconsin - Madison}
\hfill\vtop{\hbox{\bf MAD/PH/856}
            \hbox{\bf RAL-94-120 }
            \hbox{December 1994}} }
\vspace{.25in}
 \begin{title}
{\bf Minijet veto: a tool for the heavy Higgs search at the LHC }
\end{title}
\author{V.~Barger,$^1$ \ R.~J.~N.~Phillips$^2$,  \ D.~Zeppenfeld$^1$}
\begin{instit}
$^1$Department of Physics, University of Wisconsin, Madison, WI 53706\\
$^2$Rutherford Appleton Laboratory, Chilton, Didcot, Oxon, England
\end{instit}
\begin{abstract}
\begin{center} {ABSTRACT} \end{center}
The distinct color flow of the $qq\to qqH,\; H\to W^+W^-$ process leads to
suppressed radiation of soft gluons in the central region, a feature which
is not shared by major background processes like $t\bar t$ production or
$q\bar q \to W^+W^-$. For the leptonic decay of a heavy Higgs boson,
$H\to W^+W^- \to \ell^+\nu \ell^-\bar\nu$, it is shown that these backgrounds
are typically accompanied by minijet emission in the 20--40 GeV range. A
central minijet veto thus constitutes a powerful background rejection tool.
It may be regarded as a rapidity gap trigger at the semihard parton level
which should work even at high luminosities.
\end{abstract}

\newpage

\section{Introduction}

Finding ways to detect a heavy Higgs boson or longitudinal
weak boson scattering at the LHC is an issue of highest importance as
long as the nature of spontaneous electroweak symmetry breaking remains to
be established. Over the past few years considerable work has been devoted
to this topic and several techniques have been proposed to separate
the signals from large backgrounds due to QCD processes and/or the
production of $W$ bosons from the decay of top quarks. In order to identify
weak boson scattering,  $i.e.$ the electroweak subprocess $qq\to qqVV$,
tagging of at least one fast forward jet is essential~\cite{Cahn}. Early
studies~\cite{Froid,Uli,BCHP,BCHOZ} showed that double tagging is quite
costly to the signal rate because one of the two quark jets has
substantially lower median $p_T$
(order 30 GeV) than the other (order 80 GeV). Single forward jet tagging
relies only on the higher $p_T$ tag-jet and thus proves an effective
technique~\cite{BCHOZ,BCHZ,BCHSZ,DGOV}.

A study of the $WW$ signal must exploit additional identifying
characteristics. For example, the $W$ bosons from top quark decays can
be rejected by vetoing the additional central $b$ quark jets arising in
$t\to Wb$~\cite{BCHP,BCHZ}. In case of the decay
$H\to W^+W^- \to \ell^+\nu\ell^-\bar\nu$ another important discriminator is
a large transverse momentum difference between the charged
leptons~\cite{DGOV}.

In a weak boson scattering event no color is exchanged between
the initial state quarks. Color coherence between initial and final state
gluon bremsstrahlung then leads to a suppression of hadron production in the
central region, between the two tagging jet candidates of the
signal~\cite{troyan,bjgap,stelzer}. It was hoped that the
resulting rapidity gaps (large regions in pseudorapidity without
observed hadrons apart from the Higgs decay products)
could be used to select signal events. However, at LHC energies the low
signal cross sections require running at high luminosity and then overlapping
events in a single bunch crossing will likely fill a rapidity gap even if it
is present at the level of a single $pp$ collision.

In the present paper, we argue that the rapidity gap idea may be rescued at
LHC energies if we look for gaps in minijet production rather than gaps in
soft hadron production. The gluon radiation in background events is hard
enough to lead to a characteristic minijet pattern which provides an
experimentally accessible measure of the color flow in the underlying hard
event. Qualitatively, extra parton emission is
suppressed by a factor $f_s=\alpha_s {\rm ln}\; (Q^2/p_{T,{\rm min}}^2)$,
where $Q$ is the typical scale of the hard process and $p_{T,{\rm min}}$ is
the minimal transverse momentum required for a parton to qualify as a minijet.
The jet transverse momentum scale below which multiple minijet emission must
be expected is set by $f_s=1$. In the background
processes for a heavy Higgs boson the relevant hard scale may be as large as
the Higgs mass ({\it i.e.} $W^+W^-$ invariant mass). Setting $Q=1$~TeV,
$f_s = {\cal O}(1)$  may be expected for
$p_{T,{\rm min}} = {\cal O}(30\; {\rm GeV})$. Multiple minijet emission at
such a high scale should be observable even in a high luminosity environment
and therefore be useful as an event selection criterion.

As a case study to quantify these arguments, we
consider the decay mode $H\to W^+W^- \to \ell^+\nu\ell^-\bar\nu$ of a heavy
Higgs boson (typically $m_H=800$~GeV). Our first goal is to make a more
reliable estimate of the typical transverse momentum scale and the
rapidity range at which individual background events develop a high
probability for minijet activity. Second we establish that such minijets are
unlikely to be observed in signal events. Hence, a veto on these minijets
should constitute a powerful tool to isolate a heavy Higgs boson
or more generally a weak boson scattering signal. Finally, we give numerical
results for a typical search strategy at the LHC. We demonstrate that
backgrounds may be reduced well below the signal level while retaining a
sizable signal (80 events for $m_H=800$ GeV and an integrated luminosity of
$100\; {\rm fb}^{-1}$) if a minijet veto above $p_T=20$~GeV is possible.

\section{Calculational Techniques}

At least two features of soft parton emission in a hard process must be
reliably modeled in order to answer the questions raised above: i) The color
flow of the hard process and the ensuing color coherence of the soft
radiation needs to be taken into account. ii) The hard scale $Q$, which
determines the transverse momentum region where multiple minijet emission
sets in, must be determined dynamically. Both requirements are satisfied
by a full evaluation of tree level matrix elements,
including the radiation of one additional soft parton. Fortunately, Monte
Carlo programs for all the necessary signal and background simulations exist
already.

Since we are interested in heavy Higgs boson production, a simulation using
the narrow Higgs width approximation is inadequate. Instead we evaluate the
full electroweak subprocesses
\begin{equation}
qQ\quad \to\quad qQ\;\; W^+W^- + {\rm n}\;g\quad \to \quad
qQ\;\; \ell^+\nu\ell^-\bar\nu + {\rm n}\;g \;
\end{equation}
(and corresponding crossing related ones) including all $W$-bremsstrahlung
diagrams. The $W$ decays are generated in the narrow width approximation.
For the lowest order case with no gluon emission (n = 0) we use the
calculation described in Ref.~\cite{BCHZ}. In order to determine the soft
parton radiation pattern for the signal we calculate the signal cross section
for n = 1 gluon as described in Ref.~\cite{DZ}. In all cases we choose the
scale $Q$ of the structure functions and of $\alpha_s(Q^2)$ to be the
smallest individual parton transverse momentum in the final state.  For
all processes we use MRS\,A structure functions~\cite{MRSA} and we set
$\alpha_s(m_Z^2)=0.12$.

A forward tagging jet, well separated from the $W$ decay leptons, will be
part of the signal definition. Even in the top quark background such an
additional jet will almost always be produced by QCD radiation and not by
the $b$-quark arising in $t\to Wb$ decay. Hence the lowest order $t\bar t$
background is given by subprocesses like
\begin{equation}
q\bar q,\; gg \to t\bar t g,\quad {\rm with}\quad
t\bar t \to W^+W^-b\bar b \to \ell^+\nu\ell^-\bar\nu b\bar b
\; .
\end{equation}
The corresponding simulation (called $t\bar t j$ Monte Carlo in the following)
is based on the cross section formulas given in Ref.~\cite{ndeb}. When
considering the minijet activity, the top background needs to be determined
with one additional parton in the final state and we use a tree level
${\cal O}(\alpha_s^4)$ Monte Carlo program ($t\bar t jj$ Monte Carlo) which
includes the subprocesses
\begin{eqnarray}
gg       & \to & t\bar t gg\; , \\
q\bar q  & \to & t\bar t gg\; , \\
q Q      & \to & t\bar t qQ\;
\end{eqnarray}
and all crossing related ones, but which neglects Pauli interference terms
when identical quark flavors are appearing in the six quark
process~\cite{stange}. Neglecting Pauli interference is an excellent
approximation since we are interested
in the phase space region where the two final state massless partons have
very different transverse momenta and energies. In both the $t\bar t j$ and
$t\bar t jj$ Monte Carlos the top quark and $W$ decays are simulated
in the narrow width approximations. In addition, energy loss from unobserved
neutrinos in semileptonic $b$-quark decays is simulated by appropriately
decreasing the 3-momentum of the corresponding jet. In both programs the
minimal $E_T$ of the final state partons, prior to top quark
decay, is chosen as the scale of the structure functions. For the overall
strong coupling constant factors we take $\alpha_s^3 = \alpha_s(E_T(t))\,
\alpha_s(E_T(\bar t))\, \alpha_s(p_T(j))$ and $\alpha_s^4 = \alpha_s(E_T(t))\,
\alpha_s(E_T(\bar t))\, \alpha_s(p_T(j_1))\, \alpha_s(p_T(j_2))$,
respectively (where $E_T^2=p_T^2+m^2$). The top quark mass is set to
$m_t = 174$~GeV throughout.

Similar to the top quark background, the QCD $W^+W^-$ background is simulated
with $n=0$ to $n=2$ final state quarks or gluons  and is generated by a full
evaluation of all $\alpha_s^n$ tree level subprocesses~\cite{BHOZ}. Full
1-loop corrections are only known for inclusive
$W^+W^-$ production and we effectively include them by a factor $K=1.68$
for the $n=0$ process~\cite{ohnemus}. This large $K$-factor is partially due
to the emergence of new subprocesses at the $n=1$ level and therefore is not
used for the $n=1$ and $n=2$ simulations. The $W$ decays are again
treated in the narrow width approximation, the strong coupling constant
factors are taken as $\alpha_s^n = \Pi_{i=1}^n \alpha_s(p_T(j_i))$, $i.e.$
each $\alpha_s$ is evaluated at the transverse momentum of the corresponding
final state parton, and the smallest $E_T$ of the $W$'s or jets is taken as
the structure function scale.

Below we will be interested in using the higher order programs (which include
emission of soft partons) in regions of phase space where the $n+1$ jet cross
section saturates the rate for the hard process with $n$ jets. As the $p_T$
of the softest jet is lowered to values where $\sigma (n+1\; {\rm jet})
\simeq \sigma (n\; {\rm jet})$, fixed order perturbation theory breaks down and
multiple soft gluon emission (with resummation of collinear singularities into
quark and gluon structure functions, etc.) needs to be considered in a full
treatment. These refinements are beyond the scope of the present work. Instead
we employ the truncated shower approximation (TSA) to normalize the higher
order emission calculations~\cite{pps}. The tree-level $n+1\,$jet differential
cross section $d\sigma(n+1\; j)_{\rm TL}$ is replaced by
\begin{equation}\label{tsa}
d\sigma(n+1\; j)_{\rm TSA}=d\sigma(n+1\; j)_{\rm TL}
\left(1-e^{-p_{Tj,min}^2/p_{TSA}^2}\right)\;, \label{reg}
\end{equation}
with the parameter $p_{TSA}$ properly chosen to correctly reproduce the
lower order $n$~jet cross section when integrated over a given phase space
region of this hard process. Here $p_{Tj,min}$ is the smallest transverse
momentum of any of the final state massless partons. As
$p_{Tj,min}\rightarrow 0$ the final factor in Eq.~(\ref{reg}) acts as a
regulator. Note that in the case of $t\bar t$ production the top and bottom
quark transverse momenta are not included in the regularisation.

\section{Isolating the heavy Higgs boson signal}

Before discussing the minijet activity in signal and background events
we first need to define  the
event selection in terms of requirements on hard leptons and jets. The
purpose of these hard cuts is twofold: i) to reduce the various backgrounds
while keeping a large fraction of signal events and ii) to make sure that the
surviving background processes give very hard scattering events which
will have a large transverse momentum scale for additional minijet activity.

We are interested in the decay of a very heavy Higgs boson, or,
equivalently, in weak boson scattering at large center of mass energy and
in the $J=0$ partial wave. In either case the two charged $W$ decay leptons
will emerge with high transverse momentum, in the central region of the
detector, and they will be well isolated. Thus we require the presence of
two charged leptons ($\ell=e,\mu$) with
\begin{equation}\label{cut1a}
p_{T\ell}  >  50\; {\rm GeV}\;, \qquad |\eta_\ell|  <  2\; , \qquad
R_{\ell j} = \sqrt{(\eta_\ell-\eta_j)^2 + (\phi_\ell-\phi_j)^2}  >  0.7\;.
\end{equation}
Here $p_{T\ell}$ denotes the lepton transverse momentum and $\eta_\ell$
is its pseudorapidity. The $R_{\ell j}>0.7$ separation cut forbids a parton
(jet) of $p_T>20$~GeV in a cone of radius 0.7 around the lepton direction.
The lepton $p_T$ cut in Eq.~(\ref{cut1a}) is not in itself sufficient to focus
on the production of two $W$'s of large transverse momenta and large
$W$-pair invariant mass. A variable
which helps to substantially suppress $W$ bremsstrahlung backgrounds is
$\Delta p_{T\ell\ell}$, the difference of the charged lepton transverse
momentum vectors~\cite{DGOV}. We thus require
\begin{equation}\label{cut1b}
\Delta p_{T\ell\ell} = | {\bf p}_{T\ell_1}-{\bf p}_{T\ell_2}|>300\,
{\rm GeV}\;, \qquad m_{\ell\ell}> 200\, GeV\;.
\end{equation}
The additional cut on the dilepton invariant mass removes possible
backgrounds from $Z$ leptonic decays. It is largely superceded by the
the $\Delta p_{T\ell\ell}$ cut, however.

Cross sections for events satisfying the lepton acceptance criteria of
Eqs.~(\ref{cut1a},\ref{cut1b}) are listed in the first column of Table~I for
the case of a $m_H=800$~GeV Higgs boson and the
$q\bar q \to W^+W^-$ and $t\bar t$ production backgrounds. The weak boson
scattering cross section for $m_H=100$~GeV gives the electroweak background
which is still contained in the $m_H=800$~GeV line. Thus the signal cross
section is defined as $B\sigma_{\rm SIG} =
B\sigma(m_H) - B\sigma(m_H=100\; {\rm GeV})$. The 2.2~fb signal  for
$m_H=800$~GeV retains about 50\% of the total
Higgs boson signal, with the reduction largely due to the stringent
$\Delta p_{T\ell\ell}$ cut.

The $qq\to qqH$ signal is further characterized by the presence of two forward
quark jets. Typically only one of them emerges at substantial transverse
momentum and hence we use single forward jet tagging. The tagging jet
candidate is defined as the parton with the highest transverse momentum
which satisfies the general jet definition criteria
\begin{equation}\label{defjet}
p_{Tj} > 20\, {\rm GeV}\;, \qquad |\eta_j| < 4.5\;, \qquad R_{jj} >  0.7\;.
\end{equation}
Here the jet-jet separation cut is the parton level implementation of a jet
definition cone with a radius of 0.7 in the legoplot.
The tagging jet candidate is further required to fulfill
\begin{equation}\label{cut2}
p_{Tj}^{\rm tag} > 50\, {\rm GeV}\;, \qquad E_j^{\rm tag} > 500\, {\rm GeV}\;,
\qquad 1.5 <|\eta_j^{\rm tag}| < 4.5 \;.
\end{equation}
While the signal can
still be simulated at lowest order, we must include emission of an extra
parton, $i.e.$ consider $W^+W^-j$ and $t\bar t j$ production in order to
get a reliable background estimate. For the
$t\bar tj$ cross section the tagging jet is occasionally one of the
$b$-quark jets which results in a singular behavior of the cross section as
the $p_T$ of the extra parton approaches zero. This unphysical
behavior is eliminated by using the TSA (see Eq.~(\ref{tsa})) with $p_{TSA} =
20$~GeV which matches the $t\bar tj$ cross section to the $t\bar t$ cross
section within the cuts of Eqs.~(\ref{cut1a},\ref{cut1b}). As can be seen by
comparing the entries of the first two columns of table~1, requiring the
presence of a tagging jet suppresses the backgrounds by about 1 order of
magnitude while reducing the signal by 45\%. This signal reduction is mostly
due to the $p_{Tj}^{\rm tag}$ and $E_j^{\rm tag}$ requirements, which when
taken alone, account for signal losses of about 0.4~fb and 0.6~fb respectively.
Since the $p_T$ distribution of the tagging jet is relatively soft for
longitudinal weak boson scattering, one may  contemplate relaxing
the $p_{Tj}^{\rm tag}$ cut if sufficient background reduction can be achieved
by the minijet veto to be discussed later.

Another feature of the $qq\to qqH$ signal is the wide separation in
pseudorapidity of the two final state quark jets from the leptons which arise
in the Higgs boson decay. Imposing a minimal lepton tagging-jet separation,
\begin{equation}\label{cut3}
{\rm min}\; |\eta_j^{\rm tag}-\eta_\ell | > 1.7\; ,
\end{equation}
reduces the backgrounds by more than a factor 2, with little loss for the
signal (see  Table~I).

The hard cuts of Eqs.~(\ref{cut1a}--\ref{cut3}) define a trigger which selects
events like the one sketched in the legoplot of Fig.~\ref{figlego}. This
trigger is about 22\% efficient for a $m_H=800$~GeV Higgs signal while
reducing the QCD $W^+W^-$ background to an acceptable level. Top
production still drowns the signal, but can be suppressed by exploiting
the $b$-quark jet activity. In a previous study it was shown that a  veto
on the centrally produced $b$ quark jets above $p_{Tj}^{\rm veto} = 25$~GeV
is extremely effective in removing the $t\bar t$ background~\cite{BCHZ}, but
we have to be careful since at such low transverse momenta the production of
minijets via the emission of additional gluons cannot be neglected at the LHC.

\section{Minijet activity}

In order to study the minijet activity in the various processes we use the TSA
and match $\sigma (n+1\; {\rm jet})_{TSA}$ to the lower order results within
the hard cuts of Eqs.~(\ref{cut1a}--\ref{cut3}). This is achieved by setting
$p_{TSA}=63$~GeV for $\sigma (WW+2\; {\rm jet})$ and
$p_{TSA}=42$~GeV for $\sigma (t\bar t+2\; {\rm jet})$. For the weak boson
scattering process $pp \to W^+W^-3j$, slightly different values of
$p_{TSA}$ are needed to match the $m_H=100$~GeV and $m_H=800$~GeV cross
sections in the third column of table~1. Since this would lead to an incomplete
subtraction of the electroweak background when determining the signal cross
section $B\sigma_{{\rm SIG},\,TSA} = B\sigma(m_H)_{TSA} -
B\sigma(m_H=100\; {\rm GeV})_{TSA}$
we choose instead to match the $m_H=800$~GeV signal rate of 1.02 fb in Table~I
which is achieved by setting $p_{TSA}=7.3$~GeV.

The $p_{TSA}$ parameter indicates the typical scale of minijet production. We
therefore expect that  moderate $p_T$ minijet
emission is much more likely for the backgrounds than for the signal process.
The characteristic features of the additional jet activity, beyond the tagging
jet, are displayed in Fig.~\ref{figtwo}.
The pseudorapidity distributions of the jet closest to the lepton center
$\overline{\eta}=(\eta_{\ell^+}+\eta_{\ell^-})/2$ are shown in
Fig.~\ref{figtwo}(a). Here
\begin{equation}\label{defDeletalj}
\Delta\eta_{\ell j} = sign \cdot |\eta_j - \overline{\eta}| \; ,
\end{equation}
with the sign factor chosen such that the rapidity difference is counted as
positive if the second jet is on the same side of the lepton center as the
tagging jet (see Fig.~\ref{figlego}). Figure~\ref{figtwo}(a)
shows that emission of additional partons takes place in very
different angular regions for the signal as compared to the backgrounds. In a
$qq\to qqWW$ weak boson scattering event no color is being exchanged between
the two scattering quarks which emerge in the forward and backward region.
Color coherence between initial and final state radiation then leads to
suppressed emission between these two jets. Due to the large $WW$ invariant
mass the decay products of the two $W$'s emerge in the central region,
however, and thus the additional jet activity is well separated from the
charged leptons.
Indeed the signal distribution, as given by the difference between the
$m_H=800$~GeV (solid) and the $m_H=100$~GeV (dashed) curves in
Fig.~\ref{figtwo}(a), is strikingly different from that of the backgrounds.
The emission of soft gluons occurs mainly outside the
interval defined by the two quark jets~\cite{DZ}. Consequently the jet closest
to the lepton center is usually the second quark jet and not the soft gluon.
This explains the asymmetric $\Delta\eta_{\ell j}$ distribution in
Fig.~\ref{figtwo}(a): the large peak at negative values is due to the second
quark jet.  Gluon emission occuring close to the tagging jet and hence at
positive $\Delta\eta_{\ell j}$ will rarely produce the jet closest to the two
leptons.

In contrast to the signal the two background processes largely proceed by
color octet exchange between the two incident partons and color coherence
results in parton emission mainly in the central region. In $t\bar t$
production this effect is further enhanced by the $b$ decay jets
which cannot be too widely separated from the leptons since both
arise from top quark decays. A veto against this central
jet activity will clearly lead to a strong background reduction.

Figure~\ref{figtwo}(b) shows the probability to find at least one veto jet
candidate above a certain minimal transverse momentum and in the vicinity
of the leptons,
\begin{equation}\label{cutveto}
p_{Tj}^{\rm veto} > p_{T,\rm veto}\;, \qquad
\qquad \eta_j^{\rm veto} \varepsilon \;\;
[\eta_\ell^{\rm min}-1.7,\eta_j^{\rm tag}]\;\; {\rm or} \;\;
[\eta_j^{\rm tag},\eta_\ell^{\rm max}+1.7]\; ,
\end{equation}
(see shaded area of Fig.~\ref{figlego}). This probability is determined by
integrating $d\sigma_{TSA}/dp_{Tj}^{\rm veto}$ and normalizing the result to
the corresponding lower order cross section $\sigma_{LO}$ within the cuts of
Eqs.~(\ref{cut1a}--\ref{cut3}). The difference in veto probability between
the $m_H=800$~GeV signal and the two background processes is striking. In
$t\bar t$ production the veto candidate is usually one of the $b$ quarks
($\approx 80\%$ probability for $p_{T,\rm veto}=20$~GeV). In both $t\bar t$
and  QCD $W^+W^-$ production, minijet
emission due to QCD radiation sets in at much larger transverse momenta than
in the signal. This minijet $p_T$ scale is  about one to two
orders of magnitude smaller than $Q$, the momentum transfer to the color
charges which are accelerated in the hard scattering process. For the signal
$Q$ is the virtuality of the scattering weak bosons, $i.e.\; Q\approx m_W$
while the appropriate scale for the backgrounds is at least $Q=E_T(W)$ or
$Q=E_T(t)$. As a result a given probability for minijet emission is reached at
5 to 10 times larger $p_T$ scales in the backgrounds than in the signal.

In the truncated shower approximation only one soft parton is generated, with
a finite probability to be produced outside the veto region of
Eq.~(\ref{cutveto}). The veto probability will therefore never
reach 1, no matter how low a $p_{T,\rm veto}$ is allowed. At small
values of $p_{T,\rm veto}$ we thus underestimate the veto probability because
the TSA does not take into account multiple parton emission. In the soft
region gluon emission dominates and one
may assume that this soft gluon radiation approximately exponentiates. A rough
estimate of multiple emission effects is thus provided by
\begin{equation}\label{Pvetoexp}
P_{exp}(p_{T,\rm veto}) = 1-{\rm exp}\, \left[-{1 \over \sigma_{LO}}
\int_{p_{T,\rm veto}}^{\infty} dp_{Tj}^{\rm veto}\;
{d\sigma_{n+1} \over dp_{Tj}^{\rm veto}} \right]\; ,
\end{equation}
where the unregularized $n+1$~parton cross section is integrated over the
veto region of Eq.~(\ref{cutveto}) and then normalized to the lower order cross
section, $\sigma_{LO}$. For the QCD $W^+W^-jj$ background the result of this
exercise is shown as the dashed line in Fig.~\ref{figtwo}(b). It confirms the
observations made before but the deviations from the TSA result also
demonstrate the need for a quantitative calculation of the veto probability.

\section{Rapidity gaps at the minijet level}

The prime concern of a veto strategy is to retain a high acceptance of signal
events. Color coherence in the hard $qq\to qqH$ process leads to an almost
complete absence of gluon radiation between the two quark jets~\cite{bjgap,DZ}
and hence to a rapidity gap in the distribution of hadrons which result from
these soft gluons. In order to observe such a gap, however, no other sources
of soft hadrons can be allowed, either from overlapping minimum bias events in
a single bunch crossing at high luminosity or from the underlying event in a
single $pp$ collision. In the minijet model the latter is parameterized in
terms of multiple parton scattering and only a few percent (given by the
survival probability, $P_s$) of the signal events are expected to be free of
multiple interactions~\cite{bjgap,stelzer,gotsman}.  Given the small weak boson
scattering cross sections at the LHC (of order 100~fb), such a small signal
acceptance makes a ``traditional'' rapidity gap selection infeasible.

We have seen above that the different gluon radiation patterns which
are at the heart of a rapidity gap trigger become apparent in the
distributions and the rate of minijets in the 20--50~GeV transverse momentum
range. We are thus lead to define the rapidity gap trigger in terms of
minijets instead of soft hadrons. Then the survival probability of the signal
is determined as the complement to the probability that a minijet with
$p_T > p_{T,\rm veto}$ occurs
in a random bunch crossing (overlapping events) or in the underlying event
accompanying the hard scattering process. In both cases the survival
probability is given by
\begin{equation}\label{Psurvival}
P_s(p_{T,\rm veto}) = 1- {\sigma_{jj}(p_{Tj}>p_{T,\rm veto})
\over \sigma_{eff}}\; .
\end{equation}
Here $\sigma_{eff} = {\cal O}(25$~mb)~\cite{cdfdps} for minijets
produced in the underlying event and $\sigma_{eff} = [{\cal L}\;
25{\rm nsec}]^{-1} = 4$~mb for overlapping events in a single bunch crossing
at a luminosity of ${\cal L}=10^{34}{\rm cm}^{-2}{\rm sec}^{-1}$. With
single-jet cross sections of about 0.8~mb (2~mb) above $p_{Tj}=20\;(15)$~GeV
in the rapidity range of Eq.~(\ref{cutveto}) the signal acceptance
loss due to minijets in the underlying event appears to be acceptable down to
transverse momenta of order 10--15~GeV while at
${\cal L}=10^{34}{\rm cm}^{-2}{\rm sec}^{-1}$
overlapping events may produce random jets above 20~GeV $p_T$ with about 20\%
probability. This estimate agrees with the results of a more
detailed analysis of overlapping events at the LHC~\cite{ciapetta}.

In the following we shall assume that a veto on minijets with
$p_{T,\rm veto}=20$~GeV is feasible with little loss to the signal rate.
Actually, it should be possible to significantly lower this transverse
momentum cut. Central tracking may allow one to separate the $z$-vertex
position
of the hard trigger leptons from the interaction point of the minijet if
the latter arises from an overlapping event. Assuming that the LHC interaction
region will be about 10~cm long~\cite{CMS} a $z$-vertex resolution of the
charged tracks inside the minijet of a few mm should suffice. Clearly,
these questions should be addressed in experimental simulations. Here we
just want to emphasize that an elimination of minijets from overlapping events
and hence a lowering of $p_{T,\rm veto}$ would greatly enhance background
rejection with very little damage to the signal rate.

An estimate of the background reduction which can be achieved by vetoing
additional jets above $p_{T,\rm veto}=20$~GeV is demonstrated by the last
column in table~1. The minijet veto reduces the QCD $WW$ background to a
negligible level while leaving a $t\bar t$ background of about 0.5~fb.
Notice that the top production background would be a factor two larger had we
not taken into account the extra emission of soft partons in the
${\cal O}(\alpha_s^4)$
$t\bar t jj$ production process. Another measure of the background reduction is
provided in Fig.~\ref{figdelptll} where we show the distribution in the lepton
transverse momentum difference $\Delta p_{T\ell\ell}$, after our minijet veto.
A cut at $\Delta p_{T\ell\ell}=400$~GeV instead of the 300~GeV chosen in
Eq.~(\ref{cut1b}) would further reduce the background. However, trying to
make a more stringent minijet veto experimentally feasible may be the more
promising strategy.

\section{Conclusions}

The angular distribution and the typical momentum scale of the minijet
activity provide a powerful tool to distinguish the color structure of hard
scattering events. In $t$-channel color singlet exchange, such as weak boson
scattering events, there is  a suppressed central minijet activity and  the
minijets in  Higgs boson events typically carry transverse momenta well
below 20~GeV. In contrast, backgrounds such as QCD $W^+W^-$ or $t\bar t$
production involve the $t$-channel exchange of color octet gluons. This
leads to strong minijet activity at central rapidities with
$p_T\sim20$--50~GeV, which should be identifiable even in the high luminosity
environment of the LHC. Essentially, a minijet veto corresponds to a rapidity
gap search at the semihard parton level which results
in a large signal acceptance (or survival probability) even at high luminosity.

While a minijet veto appears to be a promising technique, many questions
need to be answered before its full potential as a trigger for weak boson
scattering events can be confirmed. The main question is detector related:
how low a $p_T$ threshold for the veto can be allowed without losing
significantly on signal acceptance? Since background levels would be reduced
dramatically if the veto threshold could be reduced to the 10--15~GeV range,
the search strategy for a heavy Higgs boson depends crucially on what can
finally be achieved experimentally. On the theoretical side it is necessary
to improve the predictions for the minijet activity in a region which is at
the limits of a perturbative treatment. The task is to preserve the color
coherence of multiple soft and/or collinear parton emission while keeping the
information on the momentum scale where multiple emission becomes important.
Reliable calculations
of these scales are essential to achieve a quantitative estimate of the
background reduction factors due to a minijet veto. Clearly, a leading
logarithm calculation, with an undetermined comparison scale, is insufficient.

The observation of these effects in very hard dijet events at the Tevatron
should provide an invaluable source of information~\cite{SZ}. It should
demonstrate the existence of enhanced minijet activity in hard QCD events
and give an estimate of the relevant momentum scales. Since we are dealing
with phenomena beyond the limits
of fixed order perturbation theory, a fruitful interplay between experiment
and theory appears to be the most promising way to turn minijet vetoing into a
quantitative tool to search for a very heavy Higgs boson or to make weak
boson scattering visible at future hadron colliders.

\acknowledgements
We thank A.~Stange for making his $t\bar t jj$ Monte Carlo program
available to us and we are greatful to him and J.~Ohnemus for comparisons
of numerical results.
This research was supported in part by the University of Wisconsin Research
Committee with funds granted by the Wisconsin Alumni Research Foundation and
in part by the U.~S.~Department of Energy under Contract No.~DE-AC02-76ER00881.

\newpage             

\begin{table}
\caption{Signal and background cross sections $B\sigma$ in fb after
increasingly stringent cuts. Four leptonic decay channels of the $W^+W^-$ pair
are included. The signal is defined as $\sigma(m_H)-\sigma(m_H=100$~GeV). }
\begin{tabular}{lcccc}
& lepton cuts only& + tagging jet&
{\def\arraystretch{.66}
\begin{tabular}[t]{c}
+ lepton-\\
tagging jet\\
separation
\end{tabular}}
& {\def\arraystretch{.66}
\begin{tabular}[t]{c}
+ minijet veto\\
($p_{T,\rm veto}=$\\
20~GeV)
\end{tabular}}\\
& [Eq.~(\ref{cut1a})--(\ref{cut1b})]& [Eq.~(\ref{cut2})]& [Eq.~(\ref{cut3})]&
[Eq.~(\ref{cutveto})]\\
\hline
$WW(jj)$& 27.4& 1.73& 0.57& 0.13\\
$t\bar t(jj)$& 640& 57& 25& 0.47\\
$m_H=100$ GeV& 1.18& 0.56& 0.29& 0.18\\
$m_H=800$ GeV& 3.4& 1.79& 1.31& 0.97\\[.2in]
\underline{signal}:&&&&\\
$m_H=600$ GeV&&&&0.78\\
$m_H=800$ GeV& 2.2& 1.23& 1.02& 0.79\\
$m_H=1$ TeV&&&& 0.62
\end{tabular}
\end{table}


\figure{
\label{figlego}
Legoplot sketch of a typical event after the hard cuts of
Eqs.~(\ref{cut1a}--\ref{cut3}). The shaded area represents the veto region
defined in Eq.~(\ref{cutveto}).
$\overline{\eta}$ is the average pseudorapidity of the two charged leptons.
}

\figure{
\label{figtwo}
Rapidity and transverse momentum distributions of secondary jets in
$m_H=800$~GeV ${\cal O}(\alpha_s)$ Higgs production (solid lines), $t\bar tjj$
production (dash-dotted lines) and QCD $W^+W^-jj$ production (dotted lines).
In a) $\Delta\eta_{\ell j}$ measures the pseudorapidity distance of the jet
closest to the leptons from the average lepton rapidity $\bar\eta$. Also
included is the distribution for the electroweak background as defined by the
$m_H=100$~GeV case (dashed line). The probability to find a veto jet candidate
above a transverse momentum $p_{T,\rm veto}$ in the veto region of
Eq.~(\ref{cutveto}) is shown in b). For QCD $W^+W^-jj$ production the result
for soft parton exponentiation is shown as the dashed line (see
Eq.~(\ref{Pvetoexp})).
}

\figure{
\label{figdelptll}
Dependence of signal and backgrounds on the transverse momentum difference
of the two charged leptons. In addition to the cuts of
Eqs.~(\ref{cut1a}--\ref{cut3}) a minijet veto within the veto region of
Eq.~(\ref{cutveto}) is imposed with $p_{T,\rm veto}=20$~GeV.
}

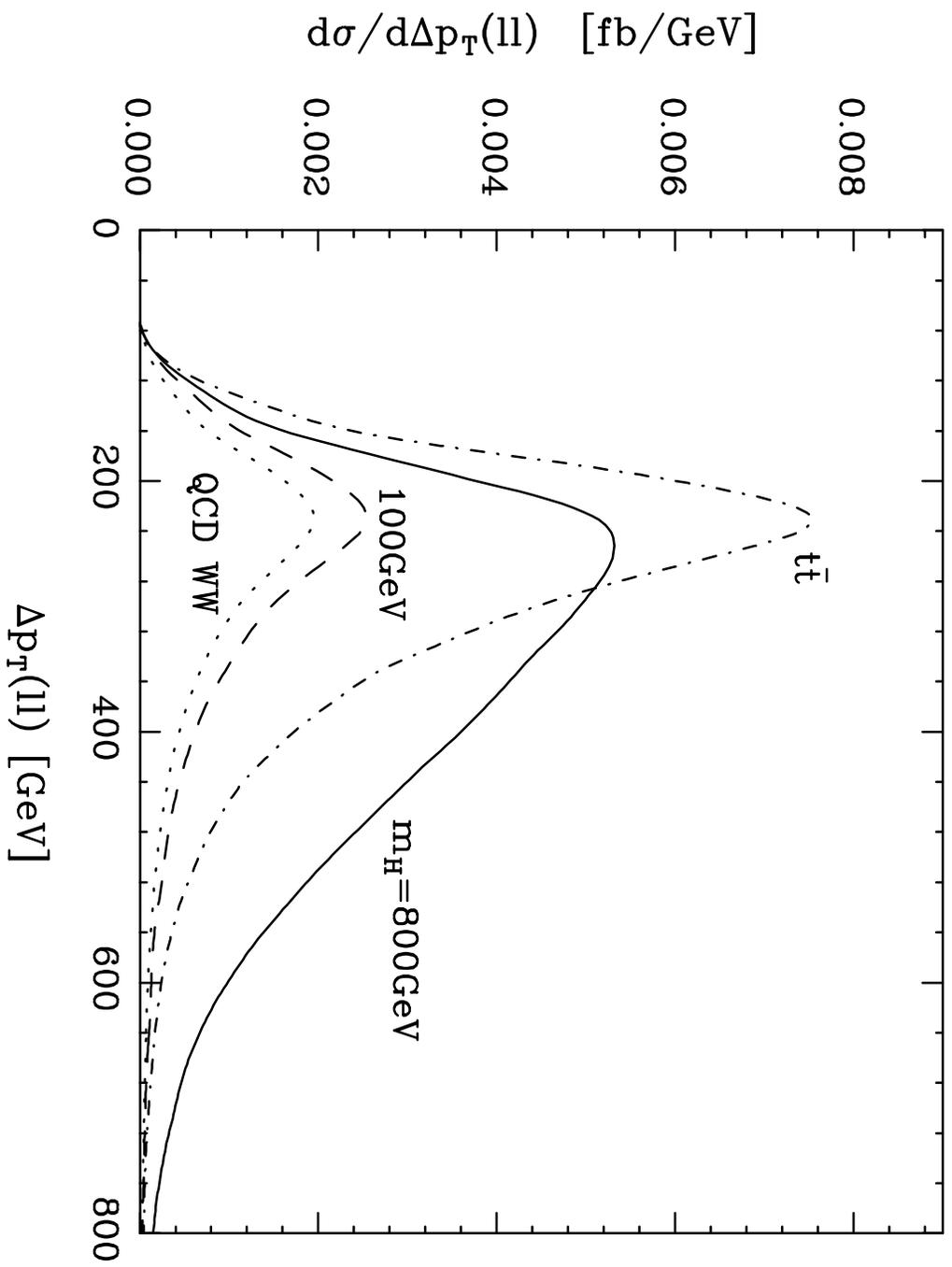
\end{document}